\documentclass[fleqn,12pt,twoside]{article}
\usepackage{espcrc1}

\usepackage{graphicx}
\usepackage[figuresright]{rotating}

\newcommand{\AmS}{{\protect\the\textfont2
  A\kern-.1667em\lower.5ex\hbox{M}\kern-.125emS}}

\title{Quantum size effects of Pb overlayers at high coverages}

\author{A. Ayuela\address[CSICDIPC]{Donostia International Physics Center (DIPC) and Unidad F\'{\i}sica de Materiales, CSIC-UPV/EHU, 20018 San Sebastian, Spain  },  E. Ogando\address[EE]{Elektrizitatea eta Elektronika Saila, Zientzia eta Teknologia
Fakultatea, UPV/EHU, 644 P.K.,48080 Bilbao, Spain}\addressmark[CSICDIPC]  and  N. Zabala\addressmark[EE]\addressmark[CSICDIPC]\thanks{Corresponding author. Address:  Elektrizitatea eta Elektronika Saila, Zientzia eta Teknologia
Fakultatea, UPV/EHU, 644 P.K.,48080 Bilbao, Spain
 Tel.:+34-94-6012538; fax: +34-94-6013071.
 \emph {E-mail address}: nerea.zabala@ehu.es}}
     
\begin{document}


{\Large Quantum size effects of Pb overlayers at high coverages}

\vspace{1cm}
A. Ayuela (1), E. Ogando (2) and N. Zabala (21*)
\vspace{.5cm}

(1) Donostia International Physics Center (DIPC) and Unidad Fıisica de Materiales
CSIC-UPV/EHU, 20018 San Sebastian, Spain

(2) Elektrizitatea eta Elektronika Saila, Zientzia eta Teknologia Fakultatea, UPV/EHU,
644 P.K.,48080 Bilbao, Spain

\vspace{1cm}
\begin{abstract}
We have studied Pb thin films  as a function of the thickness   up to 60 monolayers (MLs) using \emph{ab initio} first principles and model calculations. Magic heights corresponding to a modulated oscillatory  pattern of the energy of Pb(111) films  have been measured up to about 30 MLs. We demonstrate that  this behaviour continues even for higher thickness due to an extra second modulation pattern in the energetics of the metal film as a function of the number of atomic layers. The origin of this second modulation is the nesting of two close values of the Fermi wavelength in the (111) direction. \\

\footnotesize {Keywords:\em {Ab initio calculations, quantum wells , thin films, quantum size effects}}
\end{abstract}

\section{Introduction}
Quantum size effects (QSEs) are a key issue to understand the physical properties of nanostructures. They show up in the stability of clusters or nanowires, which is favored for some values of their radius, so-called "magic"\cite{genzken,yanson}. This behaviour has been shown to affect also the growth of islands or thin films over surfaces.  In particular these effects are very strong for Pb(111) films, which can form wide flat films of preferred thickness. Monitoring the growth  of Pb
nanoislands  over Cu(111) \cite{otero}  or Si(111)  \cite{hupalo} with
different experimental  techniques (such as He  atom scattering (HAS) \cite{Hinch},
Scanning Tunnelling Microscopy  (STM)\cite{otero}, photoemission and Surface X-ray
diffraction) has  revealed the preference  of Pb for  bi-layer growth \cite{Czoschke}.\\
The  origin of  the "magic"  height selection  is easy  to understand
qualitatively  with  a  simple  picture  of electrons  confined  in  a
potential   well    \cite{schulte}:   valence   electrons   fill
paraboloidal subbands and,  as the thickness of the  slab is increased
new quantum  well states are occupied, producing  oscillations in the
energy and related  physical properties with periodicity
$\lambda_{F}/2$  ($\lambda_{F}$ is the  Fermi wavelength of
electrons in  Pb).  When the interlayer spacing ($d$) of  Pb in
the (111) direction is considered, the approximate relation $d\approx 3\lambda_{F}/4$ gives rise to a beat pattern superimposed  to the  oscillations  \cite{edu1,Czoschke} with an even-odd change in the
magic number of atomic layers at the beats. The energy minima can be related to the magic or more stable thicknesses measured in the experiments. A quantitative description of the measured magic sizes, on the other hand, requires more sophisticated models. In some studies the support has been included to provide an accurate description of the experiments \cite{edu1,edu2,Jia}.
There are several theoretical works devoted to the study of QSEs in thin films within the density functional theory (DFT). In this  paper, we explore  larger sizes (up  to 60 MLs) of  Pb thin
films than considered in the literature by using \emph{ab initio} calculations, which include the atomic  structure of Pb films. We find a second extra modulation of the oscillations in the energetics of the thin films and explore the origin of this new modulation and its effect on the stability at higher coverages than measured up to now.

\section{Ab initio first principles calculations}
We have analized the stability of free-standing Pb(111) films as a function of the number of monolayers ($N$), up to $N$=60 ML, taking into account the atomic structure.  As  the stability is given by the energy of the  slabs, we have calculated  first the  total energy  as a  function of  
$N$.  In this study we are not interested in the quantitative agreement with the experiments, but on the amplitude and the general trend of the oscillations for thicker films than studied up to now ($N>$30). Previous jellium calculations \cite{edu1} studied the effect of the support, but in that approach only one value of the Fermi wave vector is involved, \emph{i.e.},  a spherical Fermi surface is assumed. The support can introduce a shift of these oscillations but this is not the topic of our work. The key of \emph{ab initio} calculations is that they provide a more realistic description of the Fermi surface.\\
Our calculations have been  performed with  the VASP  code \cite{vasp}  by using  the generalized
gradient     approximation     (GGA)     \cite{perdew}     for     the
exchange-correlation potential and the projector augmented-wave method
(PAW)  \cite{kresse}. Convergence of the energies versus plane-wave cutoff
(237  eV), k-mesh  (22x22),  and vacuum  (10  ML)  have been chosen  to get high  accuracy. These  values  are  in  agreement with 
reference \cite{Yu}. Previous \emph{ab initio} calculations for a small number of layers included the lattice relaxation \cite{Materzanini}, but we have considered bulk distances between atoms. This relaxation would  affect only the position of the first beat, and it is not so important for the present study. \\
To analyse the stability we have calculated the second derivative of the energy versus the number of monolayers. The results are shown in Fig. \ref{energy} with continuous line. Notice the minus sign, so that the minima correspond to stable films, as in previous jellium calculations. With the aim of comparing with the $1/D^2$ damping of the oscillations reported by jellium calculations \cite{edu3}, we have multiplied the second derivative of the energy times the square of the slab thickness. First of all we observe modulated oscillations, like "packets", with a period of about 8 ML. Between these "packets" there are beats with an even-odd slip in the magic thickness. This trend is similar to the jellium results reported in the literature. For  $N<$10 ML there are slight differences, the first beat is moved upwards. The interface properties with the substrate and the relaxation effects also can affect at these coverages, but we are not interested in that region \cite{edu1,edu2}. For larger coverages we observe a stronger damping in the  \emph{ab initio} result than the inverse square law observed for jellium. In addition, the distance between beats changes. This is more evident for $N>$30 ML, and around 40 ML the period decreases to about 6.4 ML. 
\section{Comparison with analytical models}
Jellium model calculations assume one value of the Fermi wavelength, close to 7.5 $a_0$ for Pb(111). Within that model, the energy oscillations times the square effective thickness ($D_{eff}$) of the film can be fitted to a simple sinusoidal expression which reproduces the beat pattern:

\begin{equation}
\sigma_{osc} D_{eff}^2=\sin[2k_{F} (dN+\delta_0)]
\label{eq1}
\end{equation}
where $\delta_0$ is a surface shift that accounts for the charge spill-out at the surfaces, $d$ is the interlayer spacing, and $N$ is the number of layers.\\
Our \emph{ab initio} results could not be fitted with expression (\ref{eq1}), i.e, using one value of $k_F$ in the whole thickness range, but we succeeded to fit nicely the \emph{ab initio} values with a superposition of two sinusoidal expressions for two very close values of $k_F$. We tried the following analytical description of Fig. \ref{energy}:

\begin{equation}
A_1\sin[2k_{F1}(dN+\delta_0)]+A_1\sin[2k_{F2}(dN+\delta_0)]
\label{eq2}
\end{equation}

where $k_{F1}$ and $k_{F2}$ are two close Fermi wave vectors in the (111) direction and $A_1$ and $A_2$ are their corresponding weights. The fit to such expression using $\lambda_{F1}$=7.47$a_0$, $\lambda_{F1}$=7.54$a_0$, $A_1$=0.72 and $A_2$=1.28 is really good. The result is an extra modulation over the first modulation, which gives the "packets" and first beats described by one value of $k_F$. This second modulation produces an extra second quantum beat, responsible for the stability at coverages over 30 MLs. The values of $k_F$ used for the fitting to the analytical expression (\ref{eq2}) are very close to the values reported in the literature. The second Fermi wavelength arises because the section of the Fermi surface for Pb is not circular but has a butterfly shape \cite{anderson,otero}, as shown in the inset of Fig. \ref{fourier}. In a different context, the deviation of the Fermi surface from the spherical shape has been found to affect the magnetic interlayer coupling in multilayers, being important for distances as long as 100 MLs \cite{bruno}.\\
The values of the nesting wavelengths underlying the oscillations obtained with \emph{ab initio} calculations are extracted from their Fourier transform, plotted in Fig. \ref{fourier} with continuous line. In the same plot we show the peaks corresponding to the analytical expressions that assume one (dotted line) and two (dashed line) Fermi wavevectors given by equations (\ref{eq1}) and (\ref{eq2}) respectively. We notice that both beat frequencies for the analytical model of equation (\ref{eq2}) have merged into a broad peak in the \emph{ab initio} calculations. In part this is due to the actual window of layer sizes.\\
In the inset of  Fig. \ref{fourier}, the behaviour of the oscillations with one $k_F$ (circular section of Fermi surface) is shown as compared to the two $k_F$ nesting behaviour, corresponding to the  butterfly like section of the Fermi surface. The one $k_F$ case produces a modulated oscillating structure, but the two $k_F$ plot gives a more rich pattern with a second modulation. In that sketch, two $k_F$ values with the same weight have been considered, so that the second quantum beat is clearly distinguished, and it is not accompanied by a slip from even to odd at that beat. This sketch  explains already  the origin and the trend of the \emph{ab initio} results displayed in Fig. \ref{energy}, but is not exactly the same, due to the different weights of the wave vectors.

 \section{Conclusions}
We have  shown that  there  are new  features (a second modulation structure)  in  the  quantum oscillations of Pb thin films at high coverages ($N>$30 ML), about the borderline of the thickness explored by experiments up  to now. We have obtained a good description of the energy oscillations obtained with \emph{ab initio} calculations ( $N<$60 ML) using sinusoidal expressions with two similar nesting Fermi wavelengths. The agreement is not good in the whole thickness range studied when only one value of the Fermi wavelength is considered. The second modulation of the energy oscillations would account for the magic stability of thin films at high coverages. Anyway, more experiments would be helpful at higher coverages ($N>$30 ML) in order to check this effect.

\section{ Acknowledgements}
This  work was  supported by  the  ETORTEK( NANOMAT)  program of  the
Basque government, Spanish Ministerio de Ciencia y Tecnolog\'ia (MCyT)
of Spain( Grant No. Fis 2004-06490-CO3-00) and the
European Network  of Excellence NANOQUANTA  (NM4-CT-2004-500198).  The
SGI/IZO-SGIker  UPV/EHU (supported  by  the National  Program for  the
Promotion of  Human Resources within  the National Plan  of Scientific
Research, Development and Innovation  - Fondo Social Europeo, MCyT and
Basque  Government)  is  gratefully  acknowledged  for  allocation  of
computational resources.

\pagebreak
\begin{center}

\includegraphics[scale=0.7]{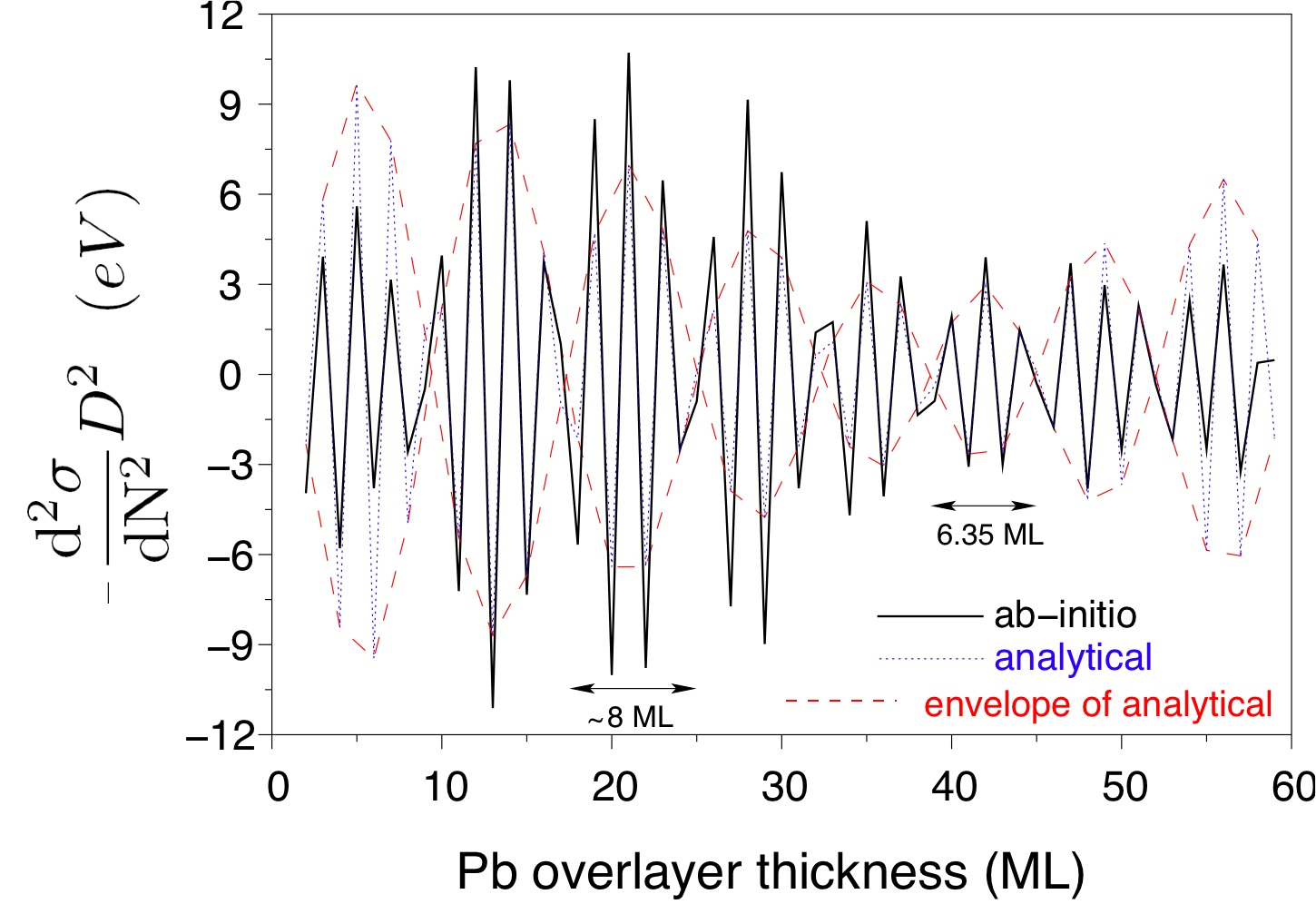}

\end{center}
\begin{figure}[htb]
\caption{Second derivative (minus sign)  of the  energy per
surface  atom times the  square of  the Pb  slab thickness  versus the
number  of  ML's  (continuous  line).  The dotted line  is  the
analytical  curve obtained  with two  values of  the  Fermi wavelength
(see the text) as given by equation \ref{eq2}. The dashed line is the envelope function of the analytical curve.}
\label{energy}
\end{figure}
\pagebreak
\begin{center}

\includegraphics[scale=0.7]{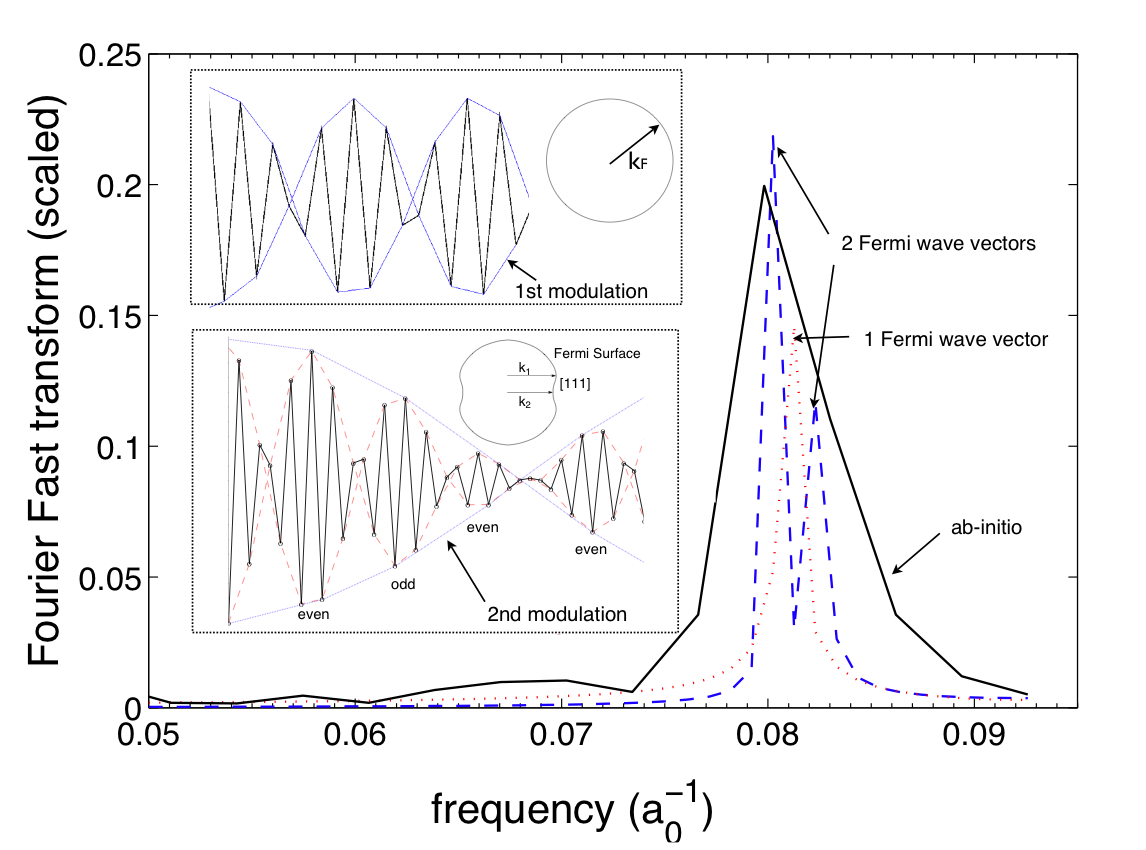}

\end{center}
\begin{figure}[htb]
\caption{ Fourier
transforms of energy oscillations. First-principle calculations  are given with full
line;  analytical expressions  for one and  two  wavelengths [equations (\ref{eq1}) and (\ref{eq2})]
are shown  with  dot and  dashed lines  respectively. In  the  plot, the
window of the analytical transforms  is three times the one of \emph{ab initio}
calculations. In the inset the oscillations corresponding to one and two Fermi wavevectors are sketched.}
\label{fourier}
\end{figure}

\end{document}